\documentstyle[preprint,aps,psfig]{revtex}
\input{psfig.sty}
\begin{document}
\title{Hadronic Entropy Enhancement and Low Density QGP}
\vskip .3cm
\author{A. Delfino $^{1}$, J. B. da Silva $^{1}$, M. Malheiro $^{1}$,
M. Chiapparini $^{2}$ and  M. E. Bracco $^{2}$ \\ 
$^{1}$ Instituto de F\'{\i}sica,
Universidade Federal Fluminense, \\ 24210-340, Niter\'oi,
RJ, Brazil.\\
$^{2}$ Instituto de F\'{\i}sica, Universidade do Estado do
Rio de Janeiro,\\
20550-900, Maracan\~a, Rio de Janeiro, RJ, Brazil.}
\vspace{.3 cm}
\date{\today}
\maketitle
\begin{abstract}
Recent studies show that for central collisions the rising of the incident
energy from AGS to RHIC decreases the value of the chemical potential
 in the Hadron-QGP phase diagram. Thus, the formation of QGP at RHIC energies 
in central collisions may be
 expected to occur at very small values of the chemical potential. 
Using many different relativistic mean-field hadronic models (RMF) at this regime  we 
show that the critical temperature for the Hadron-QGP transition is hadronic
model independent.  We have traced back the reason for this and conclude that it comes
from the fact that the QGP entropy is much larger than the hadronic entropy obtained in all the
RMF models. We also find that almost all of these models present a strong entropy enhancement 
in the hadronic sector coming from the baryonic phase transition to 
a nucleon-antinucleon plasma. This result is in agreement with the
recent data obtained in the STAR collaboration at RHIC where it was
found a rich proton-antiproton matter. 
\end{abstract}
\pacs{ }

\section{Introduction}

The understanding of nuclear matter
under extreme conditions is a crucial and indispensable aim of
nuclear and stellar physics, especially with much more experimental
information to come with RHIC and Alice/LHC accelerators.
It is now widely believed that at sufficiently high energy, 
in heavy-ion collisions,   a central hot region is formed \cite{Raf1,Greiner0,QM99,Anis}. 
This region is 
commonly associated with the existence  of a Quark-Gluon Plasma (QGP). 
Following many prescriptions, this hot region expands,
cools down and freezes out into hadrons, exhibiting a 
hadron-gas phase transition. In order to observe this transition to the 
deconfined phase, some observable signatures have been proposed as, for 
example,  electromagnetic radiation \cite{photon,photon1}, strangeness enhancement
\cite{Strange}, and J/$\psi$ suppression \cite{jpsy1,jpsy2}. High-energy heavy-ion collisions data 
has brought still more excitation about the predictions 
of QCD on this subject.  In fact, based on the results from different 
experiments done at CERN-SPS, it is believed that a new state of matter
has been formed, the quark matter \cite{QGP}.

Recently,  phase transition signatures at 
resonance-rich matter in heavy ion collisions at RHIC energies 
has been analyzed in central Au + Au collisions at $\,\sqrt{s}\,$=
200 A GeV \cite{bravina,bravina1}. These studies, using  a microscopic 
transport model (UrQMD) calculation, showed the equilibration of hot matter 
through a 
$\,T\,$x$\,\mu\,$ phase diagram. Confronting these results with RHIC, 
SPS and AGS data, it was shown that for central collisions the rising of the incident energy
implies a strong evidence for hadronic-QGP phase 
transition at very small baryon chemical potentials and temperatures 
lying between 150 to 200 MeV.  

Some time ago, using the resonance model, Heinz et al. \cite{Heinz}
, have studied the Hadron-QGP phase transition at this regime. 
They found a strong  enhancement of 
the entropy from the hadron to QGP phase. When they included 
all the resonances and mesons, 
the entropy of the hadron phase increased but a considerable  entropy gap
 still remained
between the two phases. 
In this paper we will study this transition in relativistic mean field (RMF)
 models, where we have
not only a repulsive vector interaction but also an attractive scalar
 interaction, that was not
considered in the resonance model study \cite{Heinz} .
 This scalar interaction, as we will show
in this paper, is
quite important in order to study this hadron-QGP transition at low net 
 baryon density. 
 At this regime there
is no vector field and the pressure is composed only from the nucleon thermal
 gas and from the scalar field pressures. The abrupt fall in the
nucleon mass at high temperature, seen in many of the RMF hadronic models, 
is due to the abrupt
increasing of the scalar field. This
effect resembles a phase transition when the system becomes a dilute gas of
baryons in a sea of baryon-antibaryon pairs, and it is caracterized by a large enhancement of the
entropy \cite{theis,furn0,carla}. Thus, our study in RMF models is quite important because the entropy
enhancement depends
strongly on the scalar interaction presented in these type of models. It is important to stress, that
in all these models the entropy enhancement is maximized at zero chemical potential because
when we increase the baryon density, the fall of the nucleon mass with the
temperature and also the increase of the entropy become smaller \cite{carla}.
The importance of the entropy production in  relativistic heavy ions collisions has also
been addressed  to analyze the QGP phase transition,  where  the entropy 
density jump at the phase
boundary for low net baryon density has been pointed out \cite{Greiner}.

At the hadronic level, RMF models are  able to describe well
a number of nuclear phenomena through different conceptions of
meson-nucleon and  meson-meson couplings. Most of them follows the
basic Walecka model \cite{walecka} in which improvements have been done
in order to better fit the finite nuclei data \cite{furn1,zm,zm3,nl}. 
It is known that hadronic models may present
phase transitions at higher temperatures.  For example,
in reference \cite{theis}, hot nuclear matter was
extensively studied  using the Walecka model
regarding the finite temperature behavior
for zero baryon density (nucleon-antinucleon plasma). The authors
explored this model for different scalar-vector coupling constants
to conclude that, depending on the values taken by those constants,
the Walecka model  may or may not generate a phase transition into a
nucleon-antinucleon plasma.
The order (first or second) of this phase transition depends
crucially on the scalar coupling constant of the model. 

The question we pose in this paper is whether  the 
phase transition for the hadronic sector itself, described from hadronic 
models, may or not have importance to the QGP phase transition. To
answer this question we bring to this discussion  a representative class of
 recent RMF hadronic models \cite{furn1}. 
We have also considered derivative coupling models \cite{zm,zm3} and different parametrizations of the
non-linear Walecka model such as   NL1, NL2, NL3 and NLSH models very well known in the
 literature \cite{nl}.             
Most of these models were successfully employed to calculate
nuclear matter bulk properties as well as the spectra of finite nuclei.
 The aim of our work is to present a systematic 
comparative study of a set of hadronic models, at extreme temperatures and 
very low net baryon density, and the hadronic-QGP phase transition,
emphasizing    
the entropy density behaviour in both phases. The QGP phase
is represented by a perturbative QCD derivation \cite{Heinz,boqiang}. We
studied the hadronic-QGP phase transition for a broad range of bag
constant $B$ and QCD running coupling constant $\alpha_s$ values. The particular case of 
free quarks ($\alpha_s=0$) is also explicited. This kind of phase transition has  
been investigated by many authors \cite{stoecker1,collins}. In particular,
 aspects of the role played by the nucleon exclude volume
have also been studied \cite{volume}.    

As a result, we find that practically all
the analyzed hadronic models signalize approximately the same critical
QGP temperature. This finding is quite independent whether 
the hadronic models have or not hadronic phase transition themselves 
at high temperature.  This  suggests
 strongly a RMF hadronic model independence for the QGP phase transition in this
regime. To the best of our
knowledge, this is a remarkable and unknown feature of the
hadronic models. We have traced back the reason for this and conclude 
that it comes from the fact that the QGP entropy (the slope of the pressure 
versus temperature) is much larger than the hadronic entropy obtained in 
all the hadronic models. Therefore, insofar the entropy through 
sophisticated calorimeters in the future may be a measured
quantity, we believe that at very low net baryon density,  the enhancement
of the entropy may become a signature of a QGP. Previous
works have already addressed the role played by the entropy at the QGP formation
\cite{Raf2,discontinuity,Greiner}. We have observed a large jump of the
hadronic entropy
 density, seen in almost all of the RMF models. An enhancement of the entropy per baryon
with the rising of the incident energy was also found in recent 
microscopic transport model (UrQMD) calculations \cite{bravina,bravina1}.
However, this model did not include a QGP phase but can be taken as leading credence for our hadronic
 model phase entropy enhancement signals. 
This enhancement may
  favour the hadron-QGP phase transition once  it decreases  the  latent heat 
 for the QGP formation. 
Thus, we expect  in central collisions
with a low net baryon density, the formation of a very rich baryon-antibaryon matter
 in the region of the QGP phase. This results is in agreement with recent data obtained in the
STAR collaboration at RHIC where in fact it was found a rich proton-antiproton matter
 \cite{STAR}.

The outline of this paper
is as follows. In  Sec. II we present the phase transition calculation. 
In section III we show
our results and discussion, followed by the conclusions in section IV.

\section{The phase transition calculation at high T and $\mu=0$}
\label{sec-calc}      

At low temperatures and baryon
densities (around the nuclear matter ground state), renormalization
 group arguments \cite{suryak} show that the QCD running coupling
constant  is greater than one, indicating confinement of quarks and
gluons inside
hadrons which are the appropriate degrees of freedom of the nuclear
matter in this regime. In this region, QHD
based on effective Lagrangians works quite well.
The phase transition is supposed to occur
at sufficiently high temperatures and densities, where in this
case the QCD running coupling constant becomes smaller than one,
suggesting deconfinement of quarks and gluons.
If one assumes the validity of QHD far beyond
the normal nuclear matter ground state, the transition can be
characterized
by the process of hadrons loosing  their identities, and
quarks and gluons becoming  the elementary degrees of freedom.

In order to see the hadronic-quark-gluon phase transition we
need distinct models for the two different phases of the baryonic matter.
We describe the hadronic  phase (H) by the models of Ref. \cite{furn1,zm,zm3,nl}
 in the Hartree approximation extended to
finite temperature at zero net baryon density.  For the quark-gluon plasma phase,
the equation of state is given given by \cite{Heinz,boqiang}. A study of the hadron-QGP 
phase diagram for finite baryonic density of the models presented in Ref. \cite{furn1} 
has been done in Ref. \cite{jpg}. 
 The Gibbs criteria used in the analysis of the phase transition
for zero density is given by $ \mu_{H}(T_c) =\mu_{QGP}(T_c)=0$
and $p_{H}(T_c) =p_{QGP}(T_c) $ .  

The functional, used for the hadronic phase, originates from
an effective Lagrangian, with additional physical constraints
ensuring QCD symmetries \cite{furn1}.
The  pressure is
\begin{eqnarray}
 p(M^{\ast},\rho_B)&=&-\frac{m_s^2}{g_s^2}\Phi^2
\left(\frac{1}{2}+\frac{\kappa_3}{3!}\frac{\Phi}{M}+\frac{\kappa_4}{4!}
\frac{\Phi^2}{M^2}\right)
\nonumber \\
&&-\frac{1}{2g_v^2}\left(1+\eta_1\frac{\Phi}{M}+
\frac{\eta_2}{2}\frac{\Phi^2}{M^2}+
\frac{\eta_3}{3!}\frac{\Phi^3}{M^3}\right)m_v^2W^2 \nonumber \\
&&-\frac{1}{4!g_v^2}\left(\zeta_\circ+
\zeta_{1}\frac{\Phi}{M}\right)W^4 +
W\rho_B \nonumber \\
&&+ \frac{1}{3}\frac{\gamma}{(2\pi)^{3}}\int d^{3}k\frac{k^{2}}{E^{\ast}(k)}
\; (f_{+}(k,T)+ f_{-}(k,T)) \, , \label{pfurnst}
\end{eqnarray}
where $\Phi$ and $W$ are the scalar and vector meson fields respectively
, $E^{\ast}(k)=({\bf k}^2 + M^{\ast 2})^{1/2}$ and
$\,f_{\pm} (k,T)\,$
stand for the Fermi-Dirac distribution for baryons (antibaryons).
The baryon density $\rho_B$ is defined by $            
\rho_{B} = \frac{\gamma}{(2\pi)^{3}}\int
d^{3}k \,\,\, [ f_{+}(k,T)-f_{-}(k,T)]
$. Thus, in the zero net baryon density regime we have the same
number of baryons and anti-baryons.

The constants (see Table I) are already adjusted to reproduce
some of the bulk properties of the nuclear matter ground state \cite{furn1}.
Here $W1$ stands for the usual Walecka model. $C1$ and $Q1$ are
nonlinear $\sigma$ models, having scalar field self-interaction
added to the $W1$ model in order to improve some nuclear matter
bulk properties of $W1$. Most of the
developed hadronic models has the $C1$ structure,
given by cubic and quartic scalar field self-interaction
. In this aspect it is a very representative model
category.  The scalar cubic and quartic self-interaction are usually
claimed to simulate three-body and four-body forces effects
respectively. The others as  $Q2$, $G1$ and $G2$
are models representing further improvements by including different
scalar-scalar, vector-vector and scalar-vector fields couplings.
We have also considered for the hadronic phase derivative coupling models \cite{zm,zm3} and different
parametrizations of the nonlinear Boguta-Bodmer model \cite{boguta},  NL1, NL2, NL3 and NLSH models \cite{nl}. 

For the quark-gluon
plasma phase, at high temperature, 
the pressure and the density is given by the expansion \cite{Heinz,boqiang}
\begin{eqnarray}
p_{QGP}(\mu_{q},T_{q})&=&\frac{8\pi^{2}T_{q}^{4}}{45}
\left(1-\frac{15\alpha_{s}}{4\pi}\right)
+ N_{f}\left[\frac{7\pi^{2}T_{q}^{4}}
{60} \left(1 - \frac{50\alpha_{s}}{21\pi}\right)\right.
\nonumber \\
&&+ \left. \left(\frac{\mu_{q}^{2}T_{q}^2}{2}
+\frac{\mu_{q}^{4}}{4\pi^{2}}\right)
\left(1 - \frac{2\alpha_{s}}{\pi}\right)\right] - B
\, , \label{pqgp}
\end{eqnarray}
\begin{equation}
\rho_{QGP} =  \frac{1}{3}N_{f}
\left(\mu_{q}T_{q}^2 + \frac{\mu_{q}^{3}}{\pi^{2}}\right)
\left(1 - \frac{2\alpha_{s}}{\pi}\right) \, , \label{rqgp}
\end{equation}  
where $B$ is the bag constant, $N_f$ is the number of flavors,
$\alpha_{s}$ is the QCD running coupling constant, depending on the quark-gluon plasma
temperature $T_{q}$ and the quark chemical potential $\mu_{q}$ through
the first order perturbative expression
\begin{equation}
\alpha_{s} =  4\pi\left\{ \left(11-\frac{2N_{f}}{3}\right)
\ln[ \left(0.8\mu_q^{2} + 15.622T_{q}^2\right)/\Lambda^2]\right\}^{-1}
\, . \label{aqgp}
\end{equation}                              
At the zero density regime we are interested we set $ \mu_{q}=0 $.

The model for the QGP phase which we use in
this work has some parameters such as $\Lambda$, $B$ and $N_{f}$.
As a first approximation, we examine the simplest     
case with $N_{f}=2$ (quarks $u$ and $d$ only). The QCD
scale parameter $\Lambda$ is fixed at 200 MeV, consistent
with the current data set. We will use two different
bag constants in our analysis:
$B^{1/4}=174$ MeV and $B^{1/4}=238$ MeV,
corresponding to  $B=119$ MeV fm$^{-3}$ and $B=418$ MeV fm$^{-3}$
respectively.  These are the limiting values of the broad range
of values used in the literature.   

\section{Results and Discussion}

Figure 1 shows the effective nucleon mass ratio  $m^\ast$ {\em vs.}  $T$ for the studied models. 
We see that ZM, ZM3, C1 and G2 models do not present signals of a
first order phase transition in opposite to the others.  
Then, not all of the hadronic models considered here 
 (twelve in total) have the
hot nucleon-antinucleon phase transition in the hadronic sector, manifested by
the large decrease of the effective nucleon mass $M^*$.  

 By using the Gibbs
criteria of phase equilibrium $p_{1}=p_{2}$,
$\mu_{1}=\mu_{2}=0$ and $T_{1}=T_{2}$ we have obtained the
critical temperatures for $NL1$ ( 188MeV ), $NL3$ ( 191MeV ), 
$NLSH$ ( 192MeV ), 
  $W1$ ( 186MeV ), $Q1$ ( 192MeV ), $Q2$ ( 192MeV) and $G1$ 
 ( 191MeV ) models.  The other models studied in this paper do 
not present first order hadronic phase transition. 
The critical temperatures for this  hadron nucleon-antinucleon phase transition
for hot nuclear matter at the hadronic sector itself are
remarkably close to  the critical temperature of the hadron-QGP phase transition.
This hadron nucleon-antinucleon phase transition where the effective nucleon mass becomes
very small can be seen as the Quantum Hadrodynamics version of the chiral phase transition.
It is interesting that in recent lattice calculations the chiral and the deconfinement
transitions coincide and in our calculation both transitons seem also to appear at the same
temperature depending on the value of the bag constant $B$ we choose.

We present in figure 2  the pressure for
hadronic-QGP phases (in units of the Stefan-Boltzmann pressure)  as
a function of $T$. In this figure we have used the extreme
values $B^{1/4}=174$ MeV and $B^{1/4}=238$ MeV. It is to remark how
the QGP critical temperatures are close for all the hadronic models. 
From this figure we see the critical temperatures very close to 
151 MeV for $B^{1/4}=174$ MeV  and around 198 MeV for $B^{1/4}=238$ MeV .  
The reason lies in the slope for the QGP pressure. This slope
is nothing but the entropy density, that we show  in figure 3. Notice how far,
by a factor around five ( for $\,\alpha_{s}=0\,$ ) and around three 
 ( for $\,\alpha_{s} \ne 0\,$ ), the entropy amounts are different in
both phases. It causes the hadronic model independence for the
hadronic-QGP critical temperature. 

When the hadronic model itself presents the
nucleon-antinucleon  phase transition it increases substantially
the entropy density as can be seen in figure 3. 
 However, this enhancement
is not enough when compared with the QGP entropy. This shows 
a discontinuity in the entropy 
signalizing a first order hadron-QGP phase transition, 
once we are committed with only nucleon-antinucleon in the hadronic phase. 
Let us also note that the QGP entropy density is independent of the bag
constant $\,B\,$ since it is given by ($\,{\cal E}_{QGP} + p_{QGP}
\,$)/T. Note how, in figure 2 for different bag constants, the two slopes  
 are the same for $\,p_{QGP}\,$. However, as pointed out above, 
the bag constant has effect on the hadronic-QGP value of $\,T_{c}\,$.
Therefore, if one thinks that the enhancement of the hadronic entropy
 density would help the QGP formation, only values of $\,B^{1/4}\,$ greater
than 200MeV should be considered.

 We have then included, in an {\it ad hoc} fashion, a thermal pion gas
contribution  to the hadronic phase.  By
analyzing the behaviour of ${\cal E} / p $ as a function of $T$ with
and without pions we conclude that  the change in the
critical phase transition temperature is very small.     
  When we have only nucleons and pions in the hadronic phase,
the hadronic-QGP phase transition at very low net baryon density takes place for approximately the
same value of $\,T_{c}\,$ in all the studied hadronic models.
The  hadronic-QGP phase transition is essentially given by the
 QGP phase and lies very close to the temperature where $\,p_{QGP}\,$
 crosses zero from below. 
   
The  chiral transition version we found in the hadronic models can  be
identified by a strong  decreasing in  the effective baryon mass and  an abrupt increasing of the
entropy density  in the hadronic phase ( see Fig. 1 and Fig. 3) . This large hadronic jump 
of the entropy density   favours
the hadron-QGP phase transition because it requires  a smaller latent heat at the
transition compared with the models that do not show that pure hadronic
nucleon-antinucleon plasma phase. Thus,  from our  results,  we would 
expect at very low net baryon density a formation of a very rich  baryon-antibaryon matter 
just before the QGP phase. 

It is important to stress that, even with the formation of this rich baryon plasma,
 the entropy enhancement is not enough yet to be compared with the QGP entropy density 
and still requires a large latent heat at the phase transition. 
The latent heat and entropy at the transition, as it has been discussed before on the
$\mu=0$ central rapidity region
\cite{Matsui,Matsui1,Friman}, may be seen by a large change in the specific volume that slows 
down the  time scale for the conversion of QGP into the hadron matter: a large entropy 
discontinuity  implies a long lifetime of the plasma. This point itself, the entropy
discontinuity, has deserved intense work and analysis from many authors \cite{discontinuity}. 
In fact it addresses direct the order of the hadronic-QGP phase transition. We are aware 
of the limitations of our analysis. The results are dependent on the QGP equation of state 
we are using, obtained by  a perturbative expansion in which we treat the quarks as massless. 
Note in figure 3 how the entropy density decreases with the increase of the running 
coupling constant $\alpha_s$ that at the phase transition is around $0.4$ to $0.6$ . This
suggests that a non perturbative treatment for the QGP equation of state would reduce the
entropy density. Indeed, recent lattice QCD results for  the pressure and entropy density 
\cite{Karsch,Karsch1} have also obtained a strong entropy enhancement close to the phase
transition exactly at the order of our critical hadronic entropy density (Figure 3). These
lattice results were also reproduced by  a non perturbative purely gluonic QCD calculation 
\cite{Blaizot}. Therefore, we should say that the order of the hadron-QGP phase transition is
still an open point in our work.

\section{Conclusions}

Using many different RMF models  for
the hadronic phase  and a perturbative QCD equation of state for the QGP, 
 we study the  hadron-QGP 
phase transition at zero net baryon density and high temperature.
We show that the critical temperature $T_c$ for this transition is hadronic
model independent. We have traced back the reason for this and conclude that it
comes from that the QGP entropy (the slope of the pressure versus T)
is much larger than the hadronic entropy obtained in all the RMF models.
This finding is quite independent whether the hadronic models have or not
a hadronic phase transition at high temperature when the system becomes a dilute 
gas of baryons in a sea of baryon-antibaryon pairs.

The hadronic-QGP phase transition at zero net baryonic density takes place
 for approximately the
same value of $\,T_{c}\,$ in all the studied hadronic models.
Among the studied hadronic models most of them showed a pure hadronic phase 
transition, the Quantum Hadrodynamics version of the chiral phase transition, seen by a
strong decreasing in the effective baryon masses 
and  an abrupt increasing of the entropy density in the hadronic
phase. This fact by itself does not change the hadronic-QGP value 
of $\,T_{c}\,$.  However, it is important to remark that 
the latent heat given by $\,T_{c}\,(S_{QGP} - S_{H})\, $ connecting both phases is much 
lower (but still not small, $S_{QGP}/S_{H}= 3$ with a non zero $\alpha_s$, see figure 3) 
for the hadronic models which present this hadronic phase
transition to a baryon-anti-baryon plasma, favouring the Hadron-QGP transition.
 This result enlarges the knowledge of this transition regarding RMF models 
and may reinforce different model calculations conjecturing the
enhancement of the entropy as a  signature for a hadronic-QGP 
phase transition at low net baryonic density.
This signature it is quite important nowadays with the underway experiments, such as the 
collaboration STAR at RHIC, because recent theoretical studies showed that for central
collisions the rising of the incident energy from AGS to RHIC decreases the value of 
the chemical potential in the Hadron-QGP phase diagram. Thus, the formation of QGP at 
RHIC energies may be expected to occur at very small values of the chemical potential, 
exactly the regime we are analyzing in this paper, were we expect a jump on the hadronic entropy
density due to  the formation of a rich baryon-antibaryon matter. Finally, we stress the
importance of the direct measure through calorimeters of this quantity.

\vspace{1cm}
{\bf Acknowledgements}
We  would like to express our thanks to the Conselho
Nacional de Desenvolvimento Cient\'{\i}fico e Tecnol\'ogico (CNPq) for
partial financial support.


\newpage

\begin{table}[h!]
\noindent
\caption{Model parameters, taken from Ref. [3].}
\begin{tabular}{cccccccc}
Model & $ W1$ & $ C1 $ & $ Q1$
&$ Q2 $ & $ G1 $ & $ G2 $ \\
$m_s / M$    &0.60305& 0.53874 & 0.53735 & 0.54268 & 0.53963 & 0.55410 &\\
$g_s/{4\pi}$ &0.93797& 0.77756 & 0.81024 & 0.78661 & 0.78532 & 0.83522 &\\
$g_v/{4\pi}$ &1.13652& 0.98486 & 1.02125 & 0.97202 &0.96512  & 1.01560 &\\
$\eta_1 $    &       & 0.29577 &         &         & 0.07060 & 0.64992 &\\
$\eta_2 $    &       &         &         &         & -0.96161& 0.10975 &\\
$\kappa_3$   &       & 1.6698  & 1.6582  & 1.7424  & 2.2067  & 3.2467  &\\
$\kappa_4$   &       &         &-6.6045  & -8.4836 & -10.090 & 0.63152 &\\
$\zeta_\circ $   &       &         &         & -1.7750 & 3.5249  & 2.6416
&\\
\end{tabular}
\label{table1}
\end{table}   

\newpage

\begin{figure}[t,h,b,p]
\label{fig1}
\centerline{\psfig{figure=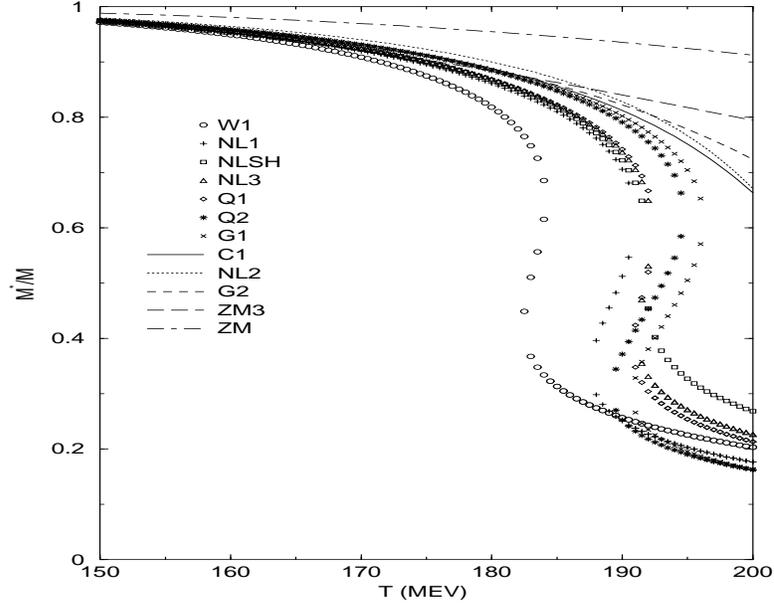,width=4.0in,height=3.2in}}
\caption{ Effective nucleon mass as a function of the
temperature
for all the hadronic models.}
\end{figure}

\begin{figure}[t,h,b,p]
\label{fig2}
\centerline{\psfig{figure=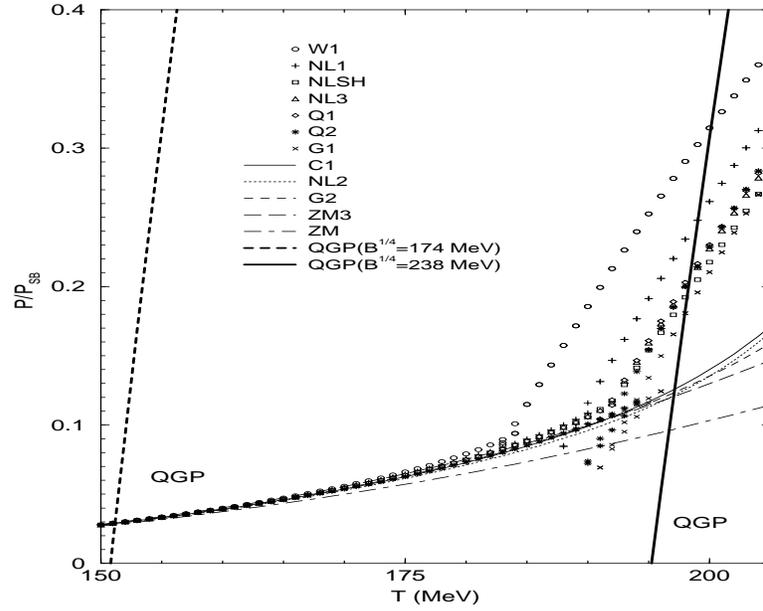,width=4.0in,height=3.2in}}
\caption{
 The pressure for
hadronic-QGP phases (in units of the Stefan-Boltzmann pressure)  as
a function of $T$.}
\end{figure}

\begin{figure}[t,h,b,p]
\label{fig3}
\centerline{\psfig{figure=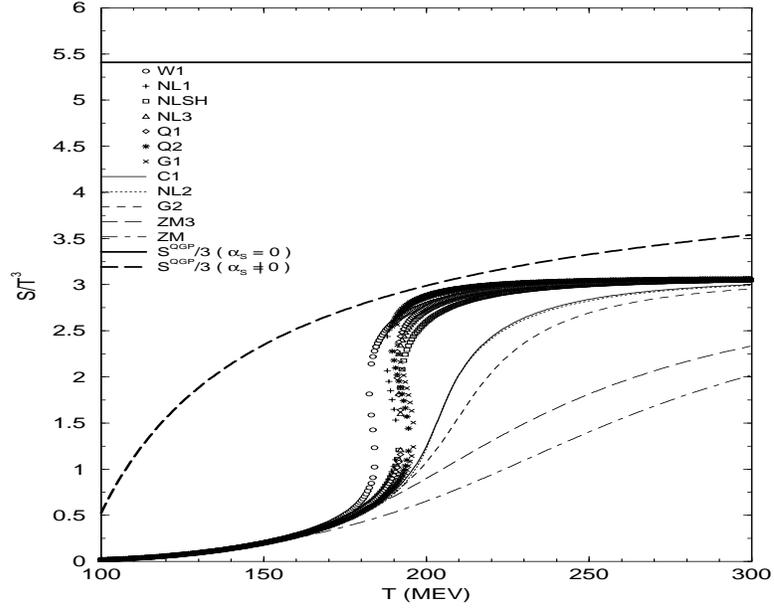,width=4.0in,height=3.2in}}
\caption{ Entropy density as a function of $T$ for all the
hadronic models and for the QGP plasma ( the QGP entropy density is
divided by 3 ) .}
\end{figure}
             
\end{document}